# Visual Political Communication in a Polarized Society: A Longitudinal Study of Brazilian Presidential Elections on Instagram

**Mathias-Felipe de-Lima-Santos**
Faculty of Humanities, University of Amsterdam, Netherlands
Digital Media and Society Observatory (DMSO), Federal University of Sao Paulo (Unifesp), Brazil
m.f.delimasantos@uva.nl/mathias.felipe@unifesp.br, https://orcid.org/0000-0001-8879-7381

**Isabella Gonçalves**
Department of Communication, Johannes Gutenberg University Mainz (JGU)
idesousa@uni-mainz.de, https://orcid.org/0000-0002-6492-9236

**Marcos G. Quiles**
Institute of Science and Technology, Federal University of São Paulo (Unifesp), Brazil
quiles@unifesp.br, https://orcid.org/0000-0001-8147-554X

**Lucia Mesquita**
Institute for Future Media and Journalism (FuJo), Dublin City University (DCU), Ireland
mesquita.lucia@gmail.com, http://orcid.org/0000-0002-2674-330X

**Wilson Ceron**
Institute of Science and Technology, Federal University of São Paulo (Unifesp), Brazil
wilson.seron@unifesp.br, https://orcid.org/0000-0002-4373-1757

**Maria Clara Couto Lorena**
Institute of Science and Technology, Federal University of São Paulo (Unifesp), Brazil
clara.couto@unifesp.br

**Abstract**

In today's digital age, images have emerged as powerful tools for politicians to engage with their voters on social media platforms. Visual content possesses a unique emotional appeal that often leads to increased user engagement. However, research on visual communication remains relatively limited, particularly in the Global South. This study aims to bridge this gap by employing a combination of computational methods and qualitative approach to investigate the visual communication strategies employed in a dataset of 11,263 Instagram posts by 19 Brazilian presidential candidates in 2018 and 2022 national elections. Through two studies, we observed consistent patterns across these candidates on their use of visual political communication. Notably, we identify a prevalence of celebratory and positively toned images. They also exhibit a strong sense of personalization, portraying candidates connected with their voters on a more emotional level. Our research also uncovers unique contextual nuances specific to the Brazilian political landscape. We note a substantial presence of screenshots from news websites and other social media platforms. Furthermore, text-edited images with portrayals emerge as a prominent feature. In light of these results, we engage in a discussion regarding the implications for the broader field of visual political communication. This article serves as a testament to the pivotal role that Instagram has played in shaping the narrative of two fiercely polarized Brazilian elections, casting a revealing light on the ever-evolving dynamics of visual political communication in the digital age. Finally, we propose avenues for future research in the realm of visual political communication.

**Introduction**

Political communication has changed with the rise of online social media platforms (SMPs), which have become indispensable tools for candidates capturing attention and gaining support from the electorate (Boulianne & Larsson, 2023; Farkas & Bene, 2021). Instagram became one of the main SMPs used by political candidates in their political communication strategy (Ekman & Widholm, 2017; Larsson, 2023). Instagram allows politicians to engage with the electorate on a more personal level, transcending the boundaries of traditional political communication (Farkas & Bene, 2021; Lalancette & Raynauld, 2019).

Despite the prevalent use of Instagram in political communication, research on this platform remains incipient. Aiming to contribute to this scholarship, this study identifies and compares the usage patterns of image-based communication on Instagram posts (*n=11,263*) by presidential candidates' public pages (*n=19*) during the 2018 and 2022 Brazilian elections.

Drawing upon a longitudinal analysis of these two Brazilian presidential contests, this exploratory study examines how Instagram has been utilized as a medium for political communication within this period. To accomplish this, our study relies on advanced computational methods, employing data science, deep learning, and computer vision algorithms to examine Instagram static images across various dimensions. Using these tools, we conducted two studies utilizing different approaches to identify patterns across these images. Our aim is to show how politicians employ visual elements to convey messages and establish their own identities. Specifically, this study seeks to investigate the utilization of Instagram's unique features and social semiotic resources to convey nuanced meanings within the realm of political campaigns within online social and cultural contexts. Using an exploratory approach, we aim to answer the following research questions (RQs) to better understand the visual politics of Brazilian electoral campaigns:

**RQ1:** What are the primary visual communication patterns employed in Instagram images during the Brazilian national elections?
**RQ2:** To what extent have the dominant visual communication patterns changed when comparing the 2018 and 2022 elections?

Our findings demonstrate the types of images that consistently dominated the posts of top-ranked candidates in both election cycles. These images tended to emphasize affection, party events, cultural particularities, and celebrations, indicating a strategic approach that resonated with the electorate. Our studies contribute to the literature by showing the main visual strategies used by political candidates on Instagram during two fiercely polarized Brazilian elections. It highlights the increasing role of visual elements in digital political communication.

**Theoretical Foundations**

*Context of Brazilian Elections: Between Polarization and Digital Campaigning*

Political context in Latin American nations differs from well-established democracies in the Global North. For instance, the democracy in the region is still emerging, and political parties have a short history and weak ideological foundations (Fiorina & Abrams, 2008; Samuels & Zucco, 2014). In Brazil, a country known for its intricate multi-party system, citizens' identification with political parties tends to be lower compared to nations with more established democratic histories and fewer political groups (Borges & Vidigal, 2018; Samuels & Zucco, 2014).

The presence of numerous parties and political coalitions with ambiguous ideological positions further contributes to the volatility of voter preferences (Mesquita Ceia, 2022). Amid these challenges, the Workers' Party (*Partido dos Trabalhadores – PT*) and the Brazilian Social Democracy Party (*Partido da Social Democracia Brasileira – PSDB*) have successfully managed to cultivate voter identification through their leadership in multiple past presidential elections, navigating a polarized and competitive political context, especially during the 2014 elections (Samuels & Zucco, 2014). However, corruption scandals changed the favoritism of both parties, and an anti-corruption agenda enabled Jair Bolsonaro, a far-right populist from the minoritarian Party Liberal Social (Partido Social Liberal – PSL), to gain significant public support, surpassing PSDB in votes in 2018. The rise of PSL and Bolsonaro was marked by increased rejection of PT (Samuels & Zucco, 2018), with Bolsonaro leveraging anti-PT rhetoric that centered on anti-corruption and anti-elite sentiments (Araújo & Prior, 2020; Davis & Straubhaar, 2020).

Especially after the election of Jair Bolsonaro, who used mainly social media channels to communicate with his voters, political parties have increasingly incorporated SMPs in their political strategy (Novoselova, 2020). The main dispute of the 2022 election took place in SMPs, and political candidates adopted different strategies to gain voters' support, ranging from negative campaigns to acclaim strategies (Borba & Vasconcellos, 2022). As part of their strategy, political actors used different SMPs to spread their messages, such as Facebook, Twitter, Instagram, Tiktok and Telegram (Lima et al., 2023). Their digital strategy responded to the increased use of SMPs, since SMPs currently had 80% of national penetration in 2022, with WhatsApp featuring in first place, followed by Instagram and Facebook (Kemp, 2022). This number represented a shift in SMPs use, as in 2018, the national penetration of SMPs was 62%, and the top 3 SMPs were Youtube, Facebook and Whatsapp (Kemp, 2018). The increased use of Instagram illustrates that political actors had to strategically think on visual messages for this SMP.

The centrality of SMPs in Brazilian political communication raised scholarly interest in understanding the political messages conveyed by Brazilian politicians. Several studies have successfully examined the political messaging during previous Brazilian elections (Lima et al., 2023; Machado et al., 2019; Recuero et al., 2021). However, one question that remains largely unexplored is the analysis of visual patterns in Instagram posts published during electoral campaigns and the differences between candidates and time periods. This article aims to bridge this gap by analyzing the images shared by Brazilian presidential candidates in both the 2018 and 2022 elections. For this purpose, we employ computational social semiotics, which offers valuable tools for examining large datasets.

*Visual Communication in Politics*

Political Communication is strongly drawn on visual elements (Veneti et al., 2019). Visual elements, such as clothing, party logos, political posters and ideological colors, have been central to communicating ideologies of political parties and social movements (Curini, 2024; Sawer, 2007). The centrality of visual elements draws back to historical social movements, such as in demonstrations of the nineteenth-century, with colors and clothing being central elements of communication of social movements' leaders (Sawer, 2007). For example, clothing was a powerful way to communicate the identity of a social movement and resonated with illiterates in more effective ways than textual messages (Navickas, 2010). After the French Revolution, particular vestments used in the social movements were perceived as dangerous by British political elites, such as the white hat, the Cap of Liberty, and the fustian jack (Navickas, 2010). The sensibilization towards particular adornments signals the power of visual elements in political communication.

In visual political communication, party logos act as a primary tool for communicating parties' identities (Skaggs, 2019), and can symbolize ideological leaning (Curini, 2024). Party logos are also persuasive because they can affect public behavior (Guilbeault et al., 2018). Often, logos use colors associated with a particular ideological leaning. The meaning of ideological leanings represented by colors have been historically constructed. At first, they were used by political and social movements' leaders, but citizens started to associate such colors with specific parties and ideologies (Curini, 2024). For example, blue signals a party's right-wing positioning, whereas red signals left-wing positioning and has been used in labor movements (Sawer, 2007). In addition, purple has been used in women's movements, whereas green has been historically associated with ecological movements (Sawer, 2007).

The increased use of SMPs in political communication strategies has further amplified the role of visual elements in messages shared by political leaders. Given the large amount of information shared in SMPs, images play a central role in capturing the attention of voters (Lilleker, 2019). In addition, images shared in SMPs have been strategically used to communicate political proposals, ideological positioning, build trust and reputation and maintain a more personalized relationship with voters (Farkas & Bene, 2021; Russman et al., 2019). Images are also important means to achieve strategic campaign goals, such as spreading negative campaigns targeting political opponents (Johansson & Holtz-Bacha, 2019) and constructing acclaim messages that strengthens a particular candidate's image (Lalancette & Raynauld, 2019). In this sense, visual SMPs, such as Instagram, have been increasingly adopted by political actors in their election campaign (Larsson, 2023). The visual nature of contemporary society has given rise to the importance of understanding how images and other visual elements communicate meaning. Yet, research on visual political communication is still incipient, particularly in the Global South. This study contributes to filling this gap by investigating visual political communication in Brazilian 2018 and 2022 elections.

*Computational Social Semiotics and Visual Politics*

Social semiotics, an interdisciplinary framework drawing upon linguistics, sociology, and cultural studies, offers a valuable lens to explore the meanings and power relations embedded in visual communication (Kress & Van Leeuwen, 2006). In political communication, visual studies examine how images, symbols, and other optic elements shape political discourses (Bleiker, 2018). As politics is inherently about power and control, the political visual aspect is crucial for understanding how politicians construct, maintain, and contest their authority on the public. Images have the potential to evoke emotions, frame issues, and mobilize support for various causes, making them an essential tool in the political communication toolbox (Grabe & Bucy, 2009; Russman e tal., 2019).

To illustrate the interplay between social semiotics and visual politics, it is necessary to analyze the political imagery, such as campaign posters, political cartoons, and photographs. These images often draw upon a range of visual resources, such as color, composition, and symbolism, to convey specific meanings and align viewers with particular ideological stances (Kress & Van Leeuwen, 2006). Those images we understand as social semiotic artifacts. Social semiotic artifacts are visual, textual, or multimodal objects, materials, or forms of communication that convey meaning and transmit messages within a social context (ibid). These artifacts are created, shared, and interpreted by individuals or groups, and they embody the cultural, historical, and social values, beliefs, and practices of those who produce and consume them. As products of social interaction, these artifacts play a crucial role in constructing and maintaining social relationships, identities, and power structures.

In the context of electoral campaigns, social semiotics examines how visual signs such as logos, colors, images, and gestures are employed to construct political messages and convey specific values and narratives. Computational semiotics concentrates on analyzing large amounts of digital data generated during electoral campaigns (Meunier, 2021). It utilizes algorithms and computational tools to identify patterns, trends, and hidden meanings in political discourse, social media messages, and other forms of online communication. Computational semiotics can reveal insights into how political messages are constructed and disseminated in the digital landscape, as well as the strategies employed to persuade voters (Meunier, 2021).

Therefore, social semiotics and computational semiotics play fundamental roles in the study and analysis of electoral campaigns. Social semiotics focuses on the examination of signs and symbols used in social communication, while computational semiotics employs computational methods and techniques to analyze and understand the meaning of signs in digital environments. The combination of social semiotics and computational semiotics in the study of electoral campaigns allows for a holistic and multidimensional approach (Meunier, 2021). Given the large amount of data examined in this study, the use of computational methods become crucial to the understanding of the images' meanings (Meunier, 2021).

**Method**

*Data Gathering and Analysis*

Our data collection process began with retrieving metadata for the 2018 and 2022 presidential candidates' public pages from various SMPs. Specifically, we utilized CrowdTangle, a social media analytics tool owned by Meta, to download comprehensive metadata for these candidates. Subsequently, we manually retrieved 11,263 publicly available static images published in public pages of political figures. Videos and GIFs were excluded from this study, focusing solely on static images.

The sample encompassed a total of 19 candidates' public pages, with nine who exclusively ran in 2018 (Geraldo Alckmin, Alvaro Dias, Cabo Daciolo, Guilherme Boulos, Henrique Meirelles, João Amoedo, João Goulart Filho, Fernando Haddad, and Marina Silva), six who ran in 2022 (Felipe D'Avila,

Leonardo Péricles, Padre Kelmon, Simone Tebet, Sofia Manzano, Soraya Thronicke), and four who ran in both elections (Ciro Gomes, Lula/Haddad, Jair Bolsonaro, and Vera Lúcia, although the latter had visual content available only for the last election). These candidates represented a broad political spectrum, spanning from far-right to far-left, and included parties that adapted their identities based on the government formation process.

We conducted two distinct studies with the collection of images associated with political figures in these elections. In both studies, we have extracted features to characterize each image, which means a set of measurements defining the image's content. Each study has employed a distinct strategy for extracting these features. In the first scenario, a deep convolutional neural network (CNN) performed the feature generation process, providing a robust feature representation, allowing the preparation of the raw images for clustering, and also laying the foundations for the subsequent analysis. However, the features generated by the CNN lack interpretability, as these features have no straightforward semantic meaning. The second study employed the Google Vision API to extract a multitude of semantic features like specific colors and objects. For instance, these semantic features allowed us to shed light on the shifting color palettes employed by political candidates in the 2018 and 2022 elections.

Throughout both phases, our analysis was complemented by a qualitative approach. Two independent coders examined the cluster contents, leveraging keywords describing the main visual themes. Their collaborative effort contributed to a nuanced analysis of political communication evolution, encompassing ideological dynamics and visual messaging strategies. Our methodological approach uncovered the strategic uses of political image usage in Brazil, ultimately fostering a deeper understanding of the visual narratives surrounding political figures and events, particularly on Instagram.

*First Study: Feature Generation with CNN*

In the initial phase, we harnessed the power of the Deep Convolutional Neural Network (CNN) MobileNetV2 (Sandler et al., 2018) to extract meaningful features from the collected images. MobileNetV2, having been pre-trained on the ImageNet dataset (Deng et al., 2009), brought the advantage of a rich foundation in recognizing various objects and scenes. For each image, we meticulously extracted a set of 1280 features from the penultimate layer of the neural network. For a detailed explanation of CNNs and the feature extraction process, see Li et al. (2022).

To enhance computational efficiency and manage the dimensionality of the feature space effectively, we leveraged Principal Component Analysis (PCA) (Ringnér, 2008). PCA, a widely-used statistical technique, facilitates the reduction of dimensionality in large datasets while preserving the maximum amount of information. This process transformed our dataset into a compact representation of 256 features while retaining approximately 82.5% of the total variance of the original data. This represents a substantial part of their variability, increasing the likelihood that the visualization will reveal meaningful patterns in the data (Hastie et al., 2009).

*Clustering Analysis*

Our exploration of the dataset's inherent structure commenced with the application of the K-means clustering algorithm. K-means, a partitioning method, segregates data points into K clusters, where K is a user-defined parameter. We conducted several experiments, each varying the number of clusters (K) to assess how different cluster numbers influenced the grouping of images.

For each K value, we followed five steps. First, we randomly initialized K cluster centroids within the feature space. Subsequently, each image's feature vector was assigned to the cluster with the nearest centroid, utilizing the Euclidean distance as the distance metric. The centroids of the clusters were then recalculated iteratively based on the mean feature vectors of the images assigned to each cluster. After that, the assignment and centroid update steps were reiterated iteratively until convergence was achieved. Convergence was recognized when either the assignments or centroids ceased to significantly change. Lastly, we calculated the Silhouette score for the resulting clustering configuration. The Silhouette score measures the similarity of an image to its own cluster (cohesion) in comparison to other clusters (separation). Higher Silhouette scores typically indicate more well-defined clusters. We evaluated distinct values of K ranging from 2 to 50. During this process, we identified three peaks (K={13, 18, 23}), aiming to identify the K value that yielded the highest Silhouette score, signifying the most appropriate number of clusters for our dataset. The Silhouette score served as a robust metric guiding our choice of the number of clusters, ensuring that our clustering results were well-founded and interpretable. In our analysis, K=23 was determined as the most suitable choice.

*Second Study: Feature Identification with Google Vision*

In the second phase of our analysis, we employed Google Vision to identify various features within the images (Omena et al., 2021). Google Vision API provides developers with a powerful tool to process and extract meaningful insights from visual data. It is a cloud-based service offered by Google that harnesses the power of computer vision, a branch of artificial intelligence focused on understanding and interpreting visual information (de-Lima-Santos & Ceron, 2021). This API enables developers to integrate advanced image analysis capabilities into their applications and services.

Key features of the Google Vision API include image analysis, optical character recognition (OCR), face detection, label detection, logo detection, landmark detection, and explicit content detection. For example, this tool can identify and label objects, scenes, and faces within images, making it invaluable for applications like image tagging, facial recognition, and content moderation. Additionally, the API can extract text from images and documents, including handwritten text, which is useful for digitization and data extraction tasks. Previous studies have used the tool to analyze the circulation of images shared on SMPs, such as Instagram and Twitter (d'Andrea & Mintz, 2019; Ferwerda & Tkalcic, 2018; Rogers, 2021).

Our focus was primarily on features pertinent to our study, including colors, labeling, face detection, logo detection, and landmark detection. This comprehensive analysis yielded a substantial number of features, which were subsequently reduced to a set of predominant features extracted from the images.

Following the feature extraction, we once again applied the K-means clustering algorithm to partition these images into clusters, ultimately identifying a total of 25 distinct clusters based on the Silhouette score.

**Triangulation Phase: Qualitative Analysis and Interpretation**

Throughout both phases of our analysis, a qualitative approach was adopted to identify and understand the clusters and their associated features. Two independent coders examined the images within each cluster and used descriptive keywords to characterize the visual content (Giesen & Roeser, 2020). By sharing individual notes with one another, the coders re-examined their analysis and reached a consensus on the best keywords describing each cluster. Subsequently, they collaborated to reconcile their findings and generate an in-depth analysis of the evolution of image usage by political figures, considering political parties and general elections. Additionally, we conducted a comparative analysis of color usage between the 2018 and 2022 elections to gain insights into shifts in visual messaging by political candidates. This qualitative analysis enriched our understanding of the clusters and their implications within the context of Brazilian politics.

**Interpretation: Qualitative Analysis**

In both studies, the qualitative analysis was fundamental for understanding the main themes used in Brazilian visual political communication. Study 1 investigated a vast array of images in the clusters emerging from this computational method. Two coders analyzed these images and employed keywords to describe the characteristics of each cluster. Subsequently, they collaborated to reconcile their findings and generate an in-depth analysis of the evolution of image usage by political figures, considering political parties and general elections. Collaborating closely, they weaved together a detailed list of themes that shed light on the evolution of image usage by political figures. By contrasting image usage in the 2018 and 2022 elections, the study aimed to unearth profound insights into the visual communication strategies and semiotics employed by the political candidates. This juxtaposition across time frames allowed them to discern trends and shifts in the way images were harnessed to convey political messages.

In contrast, Study 2, took a somewhat different approach. Rather than relying solely on human discernment, it harnessed the power of technology, specifically using Google Vision data. This AI-driven tool provided a fresh perspective by informing image features through the lens of computational analysis, assisting to identify features, such as objects, scenes, faces, and color incidences. Through the use of histograms, it sifted through vast datasets to determine the most common features within each cluster. These features were then used to systematically categorize these clusters, adding an extra layer of insight to the overall analysis. Thus, the qualitative analysis served as a bridge between the raw images and their interpretation, offering valuable context and depth to the clustering results.

The synergy between these two studies was instrumental in providing a holistic understanding of the political image usage in Brazil. Furthermore, the comparative aspect of the two studies proved to be very insightful. This comprehensive approach allowed the researchers to decode not only what was depicted in the images but also the underlying political messages and strategies. The contrast between the fully human-coded qualitative analysis in Study 1 and the mixed computational and qualitative analysis in Study 2 enriched the overall findings, showcasing the multifaceted nature of political image usage and the importance of evolving methodologies in its study.

Furthermore, the comparative aspect of the two studies proved to be very insightful. It offered a unique vantage point from which to discern the prevailing visual communication strategies and semiotics that had been adopted by political candidates over time.

**Findings**

*Study 1*

Our computational method used in Study 1 offers valuable insights into the complex nature of the clusters identified in our analysis. Figure 1 shows each image represented as a dot, with different colors indicating different clusters. The figure illustrates the intricate interconnections between these clusters, highlighting the challenge of clearly distinguishing the unique characteristics of each cluster. While the boundaries between clusters may appear blurred, our findings reveal significant patterns and relationships. This is demonstrated in Table 1 (https://osf.io/jphbw/?view_only=307c356c7e38443083d12b6a43462773), where values have been normalized to represent the percentage of images per candidate, accounting for variations in the number of images each candidate has. Our findings have unveiled valuable insights into how political candidates utilize images in their presidential campaigns in 2018 and 2022. These insights enhance our understanding of the strategies and messaging techniques employed by candidates in the ever-evolving landscape of political discourse (details in Table 2).

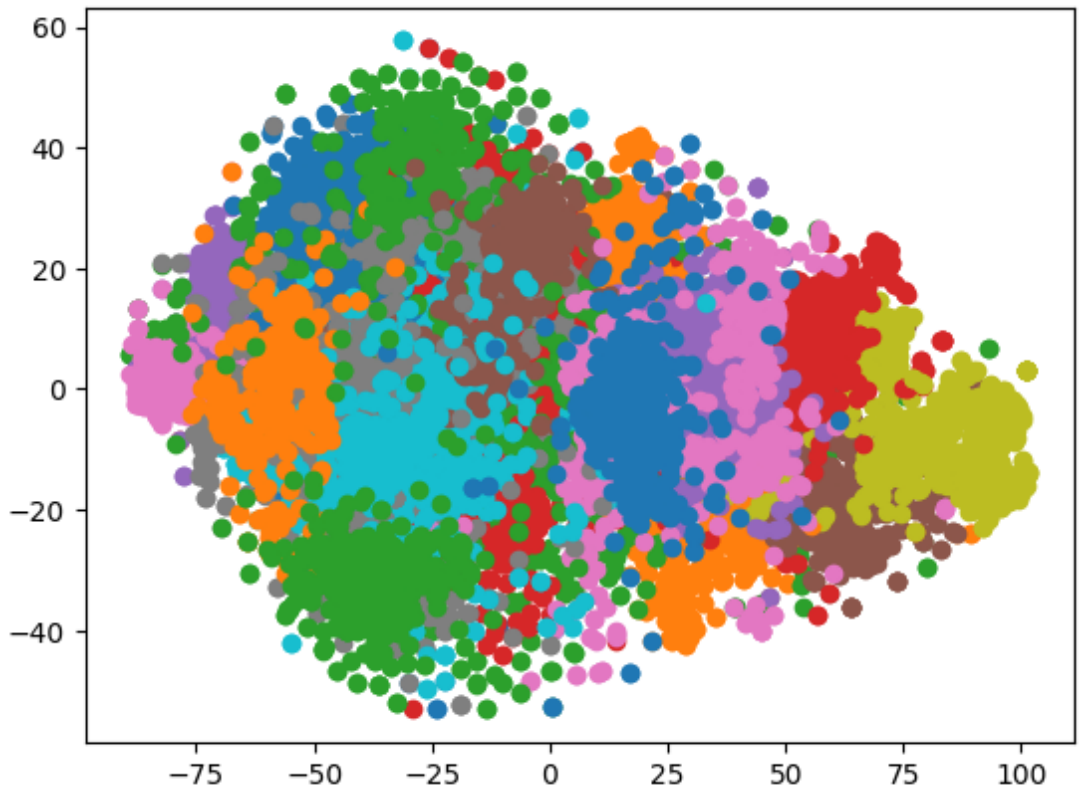

Figure 1. The chart exhibits complex interconnections of clusters and shows their silhouette.

Figure 2 provides several examples illustrating how these images were categorized based on their different features. For example, Cluster 0 has relatively low values for all political figures, suggesting that it is not strongly associated with any specific figure or entity. This cluster has relatively low differences among the candidates, with most percentages below or equal to 10%. However, it has slightly higher values for Marina Silva (Sustainability Network – REDE), Simone Tebet (Brazilian Democratic Movement –

MDB), and Jair Bolsonaro in 2022 (PL) compared to others. These candidates vary in the wider political spectrum, spanning from far-right (Bolsonaro) to left (Silva), and included parties that adapted their identities based on the government onboard (MDB). In this cluster, our qualitative analysis indicates the visual communication strategies employed by political candidates include group selfies, candidates with various groups (voters or press), and affective gestures. These images likely aim to convey a sense of connection and emotional appeal to the audience.

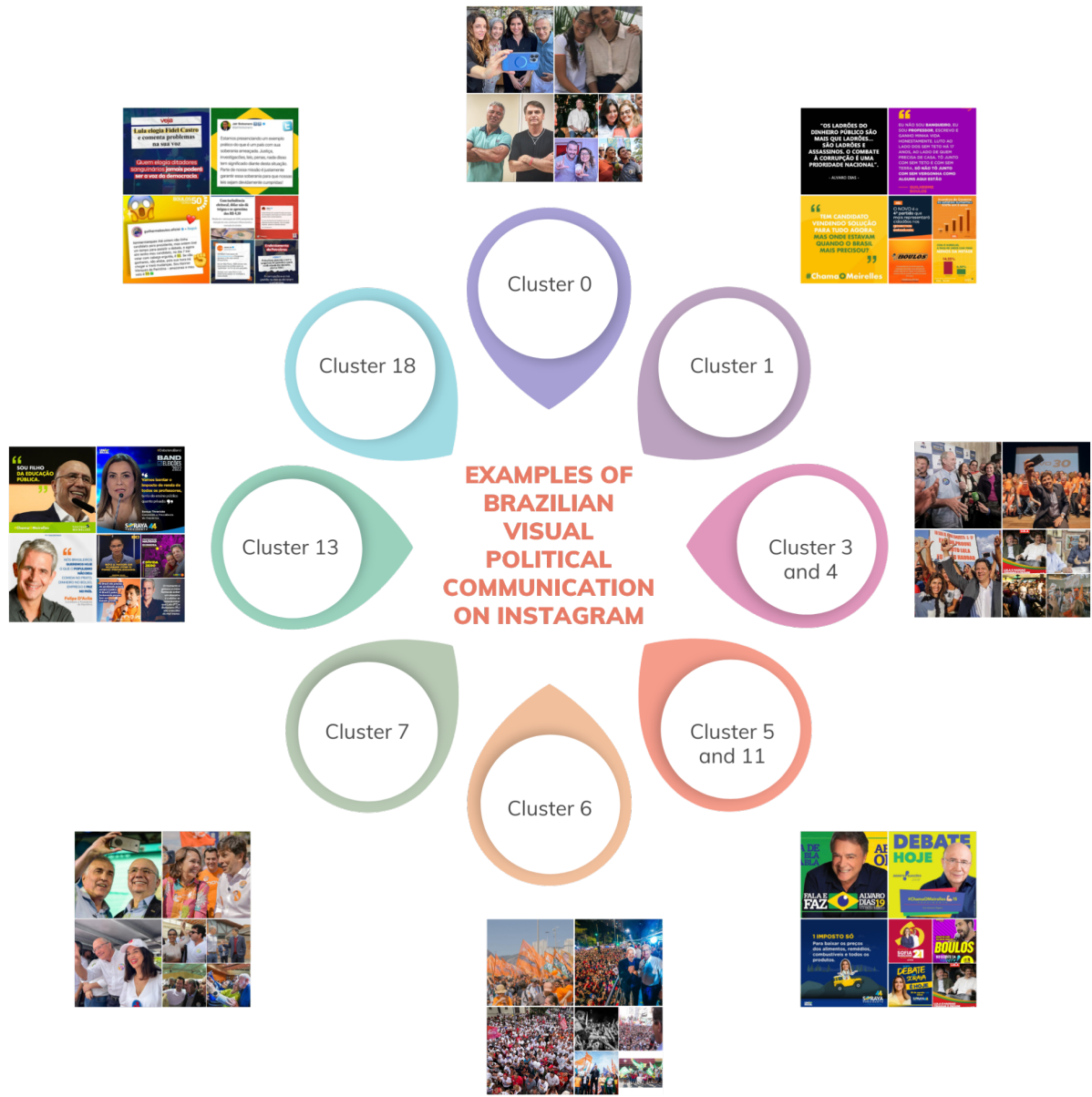

Figure 2. Examples illustrating the categorization of images based on their features in Study 1.

Similarly, Cluster 1 exhibits relatively low differences among the candidates, with all proportions below 15%. The images in this cluster are text-heavy and feature colorful backgrounds. They may include political proposals, sentences from politicians, or even negative campaign messages. It has slightly higher values for João Amoêdo (New Party – NOVO) and Álvaro Dias (Podemos Party), both of whom ran in 2018 and can be considered center-right/right parties.

Cluster 2 displays relatively low differences among the candidates, with all proportions below 9%. These images often showcase diverse groups with cultural or religious symbols such as crosses, saints, or distinctive hats. They may be taken at events with cultural significance. Two candidates, Soraya Thronicke (2022) and Álvaro Dias (2018), present higher proportions of images from the same party (Podemos). Besides them, Guilherme Boulos (Socialism and Liberty Party – PSOL), Sofia Manzano (Brazilian Communist Party – PCB), and Vera Lúcia (PSTU) have a relevant number of images in this cluster. All of them belong to left to far-left parties.

Cluster 3 primarily contains images that typically capture political party events, showing candidates accompanied by groups, along with branding materials such as logos and banners. This cluster has all proportion below 12% and is notably characterized by the content related to João Amoêdo (NOVO) and Lula/Fernando Haddad (PT), who represent opposite sides of the political spectrum. On another perspective, Cluster 4 comprises images depicting large crowd events where stages are set up, and there may also be vehicles with stages for speakers or performers. This cluster only exhibits the highest value for Ciro Gomes' 2018 campaign, indicating that he did not adopt this approach in the following election. Guilherme Boulos' (PSOL) campaign can be characterized by these two clusters.

Cluster 5 comprises images with minimal visuals and a strong focus on short texts, such as calling to vote in these politicians' numbers or to watch debates. Similarly, Cluster 11 features individual pictures accompanied by text, including quotes from interviews and TV debates. This approach is also commonly observed in Lula's campaign and was consistently used in both years, indicating its recurrent nature as a strategy. Cluster 13 also showcases some features adopted by three candidates (Soraya Thronicke, Felipe D'Avila and Henrique Meirelles) in their campaigns. It contains images with text overlaid on backgrounds. These include political proposals, expressions of thanks, calls for votes, poll results, and even negative campaign messages. This cluster has the highest value for right-wing candidates running in 2022 (Soraya Thronicke and Felipe D'Avila) and in 2018 (Henrique Meirelles).

In contrast, Cluster 6 consists of images taken from above, providing an overview of crowd events often featuring flags and the external environment. Cluster 7 includes images of small groups posing for photos. It appears that candidates may adopt various campaigning approaches without adhering to a clear standard. For instance, João Goulart Filho's visual campaign, affiliated with the Free Fatherland Party (PPL), falls within Clusters 5 and 7.

Images including screenshots of tweets and headlines from media sources can be found in Cluster 18. This has a relatively high difference of one candidate. Ciro Gomes adopted this visual approach in 2022 to share his tweets in a cross-platform communication, overcoming the limited affordances of Instagram (Bucher & Helmond, 2018; Hase et al., 2022). This was not commonly used by him in 2018, showing a different approach, which might not have been successful, as he lost votes from one election to another (Camarotto, 2022).

A comparable situation is identified in the campaigns of Bolsonaro and Lula in 2022 grouped in Cluster 20 and 21, respectively. In this year, Bolsonaro's images featured portraits accompanied by text, often related to agendas, proposals, negative campaigns, or calls for mobilization. However, this visual approach was not evident in his previous campaign. Similarly, Lula's campaign in 2022 relied on images from political party events held in external settings, often featuring candidates with supporters. This visual strategy was not commonly seen in his party's previous campaign.

These observations from the Brazilian political context indicate that these three candidates attempted to reinvent themselves during these elections, aiming to maintain or gain popularity among their voters by using visual elements that could capture their attention (Farkas & Bene, 2021). Moreover, these findings suggest that Instagram transcends the level of personalization noted by other scholars (Farkas & Bene, 2021; Lalancette & Raynauld, 2019; Larsson, 2023). In this context, the platform is utilized to communicate political proposals and ideological stances, mobilize voters through calls to action, and publicize media appearances.

In summary, this computational method generally grouped candidates with similar relevance scores, and most clusters have relatively low differences among the candidates. The specific candidates considered most relevant may vary depending on the cluster, but generally, Bolsonaro, Ciro, Lula, and their respective campaigns consistently appear as relevant candidates across multiple clusters but without a commonality between these two years.

*Study 2*

In the computational method adopted in Study 2, we have fine-tuned the model using features provided by Google Vision. This approach allowed us to gain valuable insights into the visual political communication strategies of these candidates. These insights included identifying the prominent use of certain colors in their visual content, as detailed in Table 3 (see https://osf.io/jphbw/?view_only=307c356c7e38443083d12b6a43462773). In the analysis of image clusters associated with various political candidates, it becomes apparent that different clusters exhibit varying degrees of candidate prominence (see Table 4).

Some clusters clearly highlight certain candidates, while others show relatively low differences in the candidates' presence. However, the intricate nature of the clusters identified in our analysis makes some more relevant to certain candidates than others, as shown in Figure 3. For example, Cluster 0 primarily features group images with political gestures, crowds (fans), and diverse party elements. Images in this cluster depict large gatherings of people in outdoor settings, potentially at campaign rallies or events. These images often show various gestures, such as hugging and hands up, with smiles being common among the crowd members. There is also a noticeable prominence of certain colors like orange (New Party's color), red (Workers' Party's color), and purple. Previous studies have demonstrated associations between colors and ideologies (Sawer, 2007). This association with colors was also found to a lesser extent in images shared by Lula/Haddad in 2018.

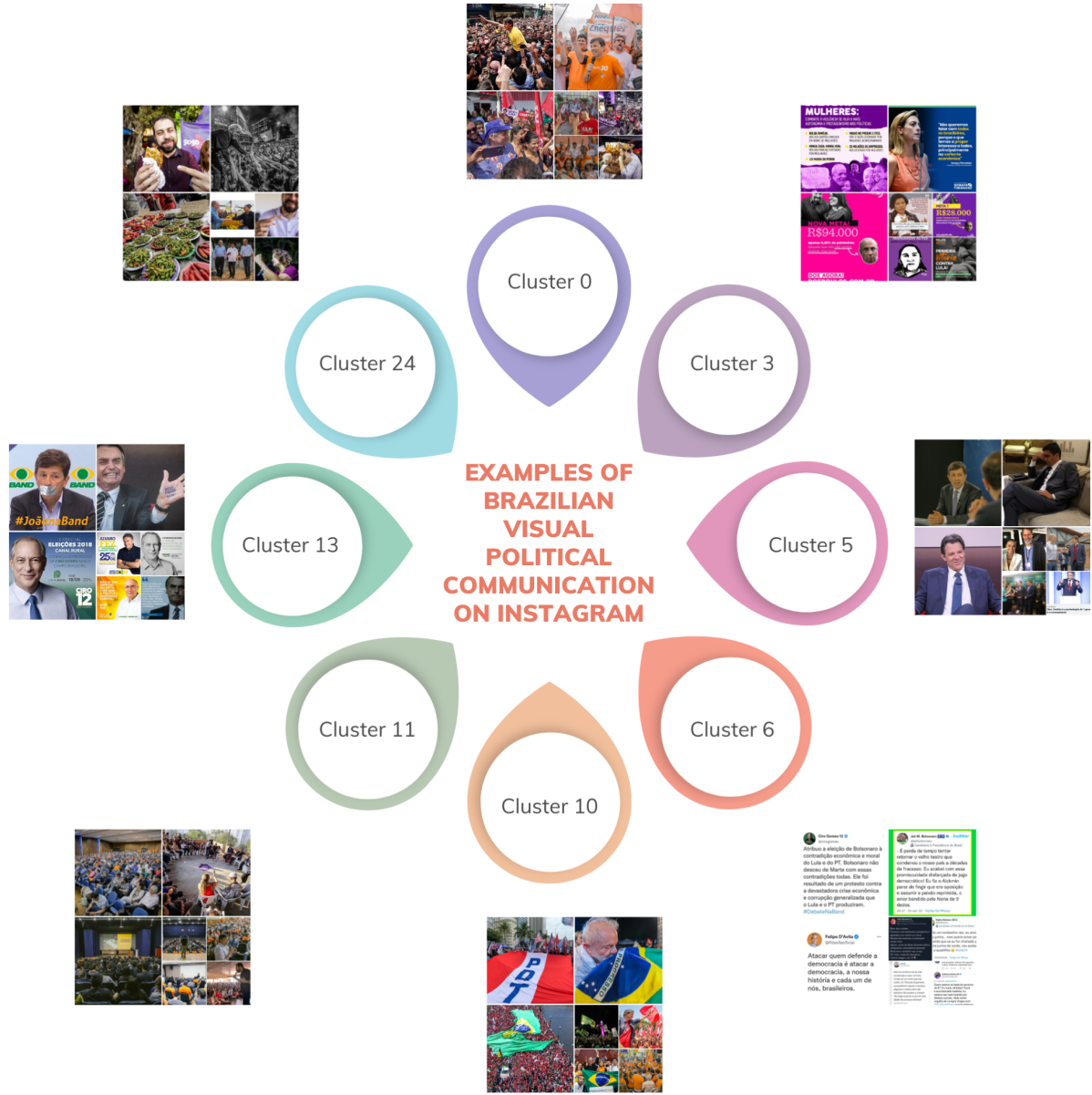

Figure 3. Examples illustrating the categorization of images based on their features in Study 2.

In the Cluster 3, candidates have a relatively low difference among them, but the proportion of images are relatively high in the most relevant ones (between 10 to 20%). Soraya Thronicke (19.41%), Vera Lúcia (14.35%), Sofia Manzano (13.90%), and Felipe D'Avila (11.77%) are the most relevant in this cluster. Images predominantly contain textual elements, including political agendas, sentences, illustrations, and sometimes negative campaign messages. Various shades of blue and purple may be used, and numerical data may be presented. Purple, often associated with the women's movement, has been predominantly utilized by female candidates like Soraya, Vera, and Sofia. This usage may suggest an intention to engage with social movements (Sawer, 2007), especially women voters in this context.

Similarly, Cluster 5 has a relatively low difference among the candidates with Cabo Daciolo (Patriota Party) as the most relevant. Most photos represent press events where individuals are formally dressed, often taking place indoors.

Cluster 6 stands out as a particularly unique and significant cluster among the analyzed political candidates. This cluster predominantly consists of images that exhibit distinctive characteristics and a remarkable concentration of certain candidates. Images in it are linked to Twitter, featuring screenshots of tweets or news articles. Some images may be related to negative campaign content, populist communication style, or self-acclaim. Ciro Gomes (29.84%) dominates this cluster with a substantial presence, making him the most relevant candidate within these images. Felipe D'Avila (17.95%), despite not being as prominent as Ciro Gomes, still has a substantial representation within these images. They are followed by Bolsonaro, who used this approach in his two presidential campaigns. This is in line with previous findings indicating that politicians use Instagram to reinforce their personal image. (Lalancette & Raynauld, 2019).

Images combining portraits with quotes and dates indicating scheduled media appearances are also commonly found in Cluster 13, with a substantial presence of Jair Bolsonaro in 2022. In these images, candidates are often seen wearing clothes with collars.

Cluster 24 consists of images of groups interacting with gestures and visiting various sites, often associated with food-related activities. This cluster is well distributed among the candidates, with Padre Kelmon, João Goulart Filho, and Jair Bolsonaro (in both years) being the most relevant candidates in this regard. According to prior findings, these food environments may communicate a "populist" perspective by incorporating elements associated with the common people (Demuru, 2021). In the Brazilian context, populism is closely linked to anti-elite sentiments (Araújo & Prior, 2021). It is likely for this reason that Jair Bolsonaro and Padre Kelmon, two populist candidates, are prominent in this cluster.

The results of this method unveiled intriguing patterns and insights regarding how these candidates utilized visual content to engage with their audiences. It identified distinct clusters, some of them revealing a unique facet of political messaging, offering a comprehensive view of the visual political communication strategies employed by various candidates. First outdoor gatherings with vibrant colors were a common theme, with candidates from different political spectrum like João Amoêdo and Lula featuring prominently. This suggests a strategy of portraying inclusivity and positivity in campaign imagery.

Additionally, formal indoor press events appeared as another significant context, with Cabo Daciolo emerging as a notable candidate in this setting. Navickas (2010) discusses how clothing can communicate ideology. For example, Cabo Daciolo, an extremely conservative and devoutly religious candidate, may have chosen formal clothes as a means of communicating his conservatism, seeking resonance with a religious audience.

The study also highlighted the prominence of Twitter in campaign visuals, with Ciro Gomes as a dominant figure. This suggests a recognition of the platform's importance in shaping public opinion and disseminating campaign messages. Moreover, the use of images combining portraits with quotes and dates was evident, particularly in the case of Jair Bolsonaro in 2022. This approach aimed to convey specific messages and scheduled appearances to the audience. Lastly, images depicting group activities with smiles, gestures, and expressions of happiness are a common theme among candidates such as Padre Kelmon, João Goulart Filho, and Jair Bolsonaro (in both elections). This approach likely aimed to establish a connection with voters through camaraderie and communal experiences. It also shows the role of visuals in constructing personalization (Larsson, 2023; Farkas & Bene, 2021). Although these three politicians are prominent, Cluster 24 has a relatively low difference among the candidates, indicating that this is a common approach among all candidates.

In summary, this method identified image clusters that represent a diverse range of political visual content, from crowd gatherings and political speeches to behind-the-scenes moments, social media content, and interactions with voters and constituents.

*The Differences of Political Visual Content: A Comparison of Two Presidential Elections*

Our Study 1 identified a few features that were shared by the same candidates or parties across the two elections. However, Study 2 allowed us to identify some features that were present throughout the entire dataset and were visible across various candidates.

Notably, images portraying group activities, often characterized by the presence of smiles, friendly gestures, and expressions of happiness, were frequently disseminated by all candidates, with a particular emphasis on right-wing candidates like Padre Kelmon, João Goulart Filho, and Jair Bolsonaro (in both campaigns). These images serve as visual representations of camaraderie and convey a sense of unity and shared enthusiasm within their respective campaigns. The deliberate use of such imagery suggests an effort to connect with the electorate on an emotional level, creating a positive and relatable image of the candidates. This approach underscores their desire to resonate with voters by showcasing moments of collective joy and togetherness, which can be a powerful tool in political communication. Additionally, these visuals may humanize the candidates, making them appear approachable and in touch with the aspirations and happiness of the people they seek to represent. This is in contrast to the traditional political figure, who often appears more formal and serious. This casual use of visuals can be seen as a way to communicate anti-elite sentiments and establish a deeper connection with the people (Demuru, 2021). In summary, the prevalence of these images highlights the strategic importance of projecting a positive and harmonious image in political campaigns.

Comparable strategies were evident in both Jair Bolsonaro's campaigns. His visual communication primarily featured images rich in textual elements, encompassing political agendas, phrases, and occasionally, critical campaign messages. This textual emphasis suggests a deliberate attempt to convey specific political messages and positions directly to the audience. Furthermore, Bolsonaro's use of Twitter-related images, such as screenshots and news articles, aligned with a narrative characterized by negative campaign content, a populist communication style, and an element of self-promotion. In fact, a prior study has discussed the use of images for negative campaigning (Johansson & Holtz-Bacha, 2019). This approach reflects his utilization of social media as a platform for directly engaging with the public, often employing confrontational and attention-grabbing tactics adopted by populist leaders (Bene et al., 2022). Moreover, on his Instagram account, Bolsonaro frequently shared images that combined portraits with quotes and dates, often signaling forthcoming media appearances. The consistent portrayal of himself in suits or formal shirts in the majority of these posts communicates a sense of seriousness and commitment to his political mission. This attire choice is a visual cue aimed at projecting a professional and dedicated image to his followers and potential supporters. In essence, Bolsonaro's visual campaign strategy across these two elections was marked by a common combination of social media engagement, negative messaging, and a deliberate use of imagery to communicate his political persona and agenda.

In contrast, Ciro Gomes employed distinct strategies during his two separate election campaigns. In the 2018 campaign, his visual communication often featured images that combined portraits with quotes and dates, frequently serving as markers for scheduled media appearances. Additionally, his campaign imagery included pictures of individuals engaged in public speaking, often at party events. These visuals encapsulated speech acts, direct interactions with voters, and settings characterized by outdoor environments and spontaneous moments. This approach in 2018 conveyed a strong emphasis on Gomes' accessibility, his direct engagement with the public, and his commitment to scheduled media engagements. These individual image choices serve to reinforce the candidate's image as competent and demonstrate the use of Instagram to bolster a politician's professional image (Lalancette & Raynauld, 2019).

However, the 2022 campaign saw a noticeable shift in Ciro Gomes' visual messaging. Twitter and news article screenshots took center stage, reflecting a different tone and focus. These images were predominantly associated with negative campaign messages, a populist style, and an element of self-promotion. The prominence of these visuals in 2022 signaled Gomes' utilization of social media and news coverage to engage with the public in a more confrontational and assertive manner. In summary, Ciro Gomes' visual campaign strategies completely changed between the two election years, with 2018 highlighting accessibility and direct engagement, while 2022 emphasized a more combative and populist style of communication through social media and news-related visuals.

Lula's visual campaign in 2022 exhibited a notable emphasis on certain aspects that had already made appearances, albeit to a lesser extent, in his 2018 campaign. In particular, a recurring theme in 2022 was the prevalence of images depicting large gatherings of people in outdoor settings, possibly indicative of campaign rallies or events. These visuals often captured a variety of gestures, with the prominent use of Party color (red). As mentioned before, colors have an important role in communicating ideology (Sawer, 2007). Additionally, smiles among the crowd members were a common feature, conveying a sense of positivity and enthusiasm within his campaign. In this most recent election, Lula's imagery was more closely associated with formal political party events and televised appearances, frequently showcasing individuals dressed in suits. Moreover, there was a heightened presence of external party events marked by the display of flags—be it party flags, social movement flags, or national flags. Similar to colors, the use of adornments serve to communicate ideology (Navickas, 2010). These visuals also conveyed a strong sense of political identity and solidarity, emphasizing the collective aspect of his campaign.

Furthermore, images portraying group activities, often characterized by the presence of smiles, friendly gestures, and expressions of happiness, remained a consistent feature across both campaigns. These visuals underscored Lula's commitment to projecting a sense of unity and shared enthusiasm among his supporters, creating an engaging and positive image that transcended both election years.

Other political parties, which ran with different candidates for these two elections, demonstrated a notable change in their campaign approaches between one campaign and another. Some parties chose to strongly emphasize certain elements from their 2018 campaigns, while others opted for a complete overhaul in 2022.

For certain parties, the continuity of specific aspects from their 2018 campaigns into 2022 was particularly evident, such as Podemos. These parties saw value in maintaining and building upon certain successful strategies, resulting in a coherent visual narrative across election cycles. This approach may be aimed to capitalize on the familiarity and resonance of established campaign themes and imagery.

Conversely, other parties chose a stark departure from their 2018 campaign strategies in 2022, such as NOVO and MDB. This shift signaled a deliberate effort to adapt to changing political dynamics, voter preferences, or evolving communication trends, aiming to secure votes from a polarized election. By reinventing their visual campaigns, these parties sought to present a fresh image to the electorate, one that aligned more closely with the contemporary political landscape. In essence, the divergent approaches taken by parties in their campaign visuals underscored the dynamic nature of visual political messaging.

*The Entanglements Between the Study 1 and Study 2*

Study 1 and Study 2 complement each other in their findings, offering different perspectives on candidates' visual political communication. Study 1 primarily revealed a limited degree of differentiation among the candidates, whereas Study 2 brought to light additional variations in their visual strategies. To assess the similarities and distinctions between these studies, we employed the Jaccard index, also referred to as the Jaccard similarity coefficient. This statistical measure is used to quantify the likeness and dissimilarity between sample sets, helping us identify commonalities (Vogler et al., 2020).

In Figure 4, the intersection between these two sample sets is highlighted in yellow, illustrating that a few clusters were shared between the two studies. However, a more detailed examination reveals that some clusters displayed specific similarities, particularly those featuring screenshots of tweets and news stories. This observation underscores the nuanced nature of candidates' visual communication strategies. Some elements remained consistent across the two studies, such as the utilization of Twitter-related visuals. However, both studies present notable disparities.

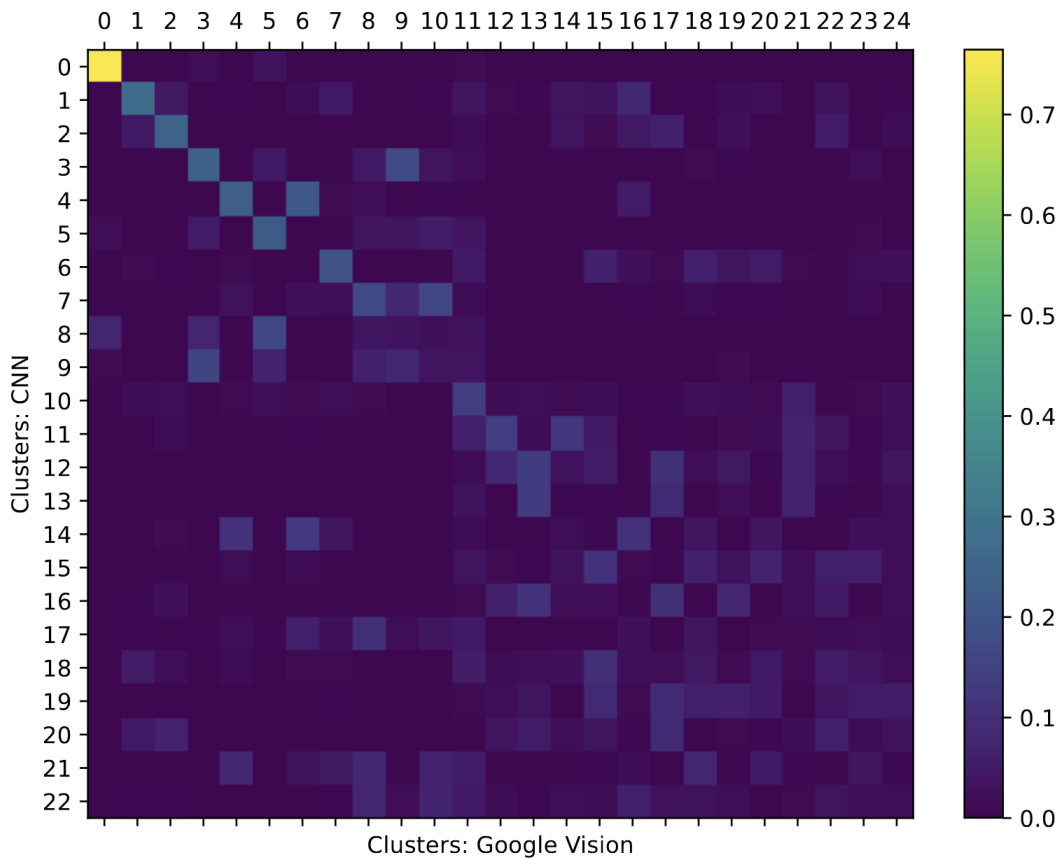

Figure 4. Correlation matrix between both Study 1 and 2

**Discussion and Conclusion**

Images play a critical role in contemporary political communication by serving as powerful tools for generating engagement and connecting with citizens. In this regard, platforms like Instagram have emerged as indispensable arenas for political strategies in various countries (Larsson, 2023). Through different images' choices, political candidates can capture the attention of voters and efficiently communicate their messages. The prevalent use of images in today's political communication derives from their unique ability to transcend language barriers, convey complex messages, and evoke emotional responses, all of which are crucial elements in mobilizing citizens to participate in the democratic process (Grabe & Bucy, 2009). As such, the impact of images in shaping political discourse cannot be underestimated, making them a vital component of contemporary political communication strategies (Ekman & Widholm, 2017; Farkas & Bene, 2021).

The protagonist role of images and Instagram in current electoral campaign strategies requires a combination of methods to examine patterns and trends among and between candidates, as well as through a longitudinal lens. This article employed two studies, the first using feature generation via CNN and the later Google Vision API. The use of two distinct computational methods allowed the examination of larger samples, as well as the confirmation of important patterns across these images. To enrich our analysis, we opted for qualitative coding the images after their separation in clusters. The combination of computation-human approach contributed to the understanding of particularities that algorithms alone are unable to uncover (Andreotta et al., 2019; Ceron et al., 2021; Karjus, 2023). Findings from Study 1 and Study 2 provide valuable insights into candidates' visual political communication strategies. These insights enhance our understanding of how candidates utilized images in their presidential campaigns in both 2018 and 2022.

Study 1 offered a comprehensive analysis of the visual content employed by candidates in the two election cycles. It became evident that some clusters exhibited a blurred boundary between them. For instance, group images with political gestures and diverse cultural elements were employed by candidates from various parts of the political spectrum. This is also evident in images characterized by a stronger voter contact, featuring party celebrations, group selfies and affective gestures. Moreover, another attribute found among different strategies is the use of Brazilian cultural richness, with images showcasing regional traditions and religious celebrations. These are in line with previous studies examining the use of Instagram in international campaigns, which indicate that Instagram is often used to foster a sense of unity, with images focusing on more personalized and humanized content, such as celebrations and more positive tone (Farkas & Bene, 2021; Towner & Muñoz, 2018).

In Study 2, the utilization of fine-tuned models and features from Google Vision added depth to our understanding of candidates' visual communication. Our results revealed that candidates continued to employ a diverse range of visual elements. These ranged from images emphasizing emotional connections and group dynamics to formal indoor settings, press events, and the use of social media, such as Twitter, to engage with the public. The importance of self-promotion and conveying political messages directly to the audience through text-heavy visuals and screenshots of tweets or news articles became evident.

However, we also find a large quantity of images featuring less personalized content. Text-heavy, edited images focusing on proposals, candidate's sentences in TV debates, and negative campaign messages could be commonly found among candidates. These findings differ from results in other contexts, in which Instagram is more strongly used for more spontaneous and casual types of images (Farkas & Bene, 2021).

Comparing the two studies, we find a blend of continuity and change in candidates' visual strategies. While some elements remained consistent, such as the use of Twitter-related visuals, candidates also adapted their approaches to suit the evolving political landscape and communication trends. This article shows the evident use of colors to connect voters with the parties' ideology as well as these images present the use of political adornments, such as buttons and flags (Sawer, 2007). Similarly, the dichotomy between formal and casual clothing is also presented in these images (Navickas, 2010). All these visual elements contribute to understanding the strategies adopted and the type of message politicians aim to convey to their electors (Curini, 2024).

Unlike other studies that emphasize the personalization aspect of Instagram (Lalancette & Raynauld, 2019; Larsson 2023), our research demonstrates that Instagram serves various purposes in the Brazilian context and has a broader function. The platform is used to communicate political proposals and ideological stances, mobilize voters through calls to action, and publicize media appearances.

Lastly, our finding demonstrates that during the Brazilian elections, Instagram is also used for the type of content that is usually more popular in Twitter and Facebook platforms in other contexts (Boulianne & Larsson, 2023), overcoming the limited affordances of Instagram and adopting a cross-platform communication (Pearce et al., 2020). One question that remains open is to what extent the use of less-personalized content in Instagram is popular among voters in the Brazilian context. Future cross-platform studies should help uncover the types of content generating more engagement.

Our research has demonstrated the visual trends within the Brazilian political communication landscape, making a valuable contribution to the relatively incipient field of research into the visual affordances of electoral campaigns. Employing advanced computational methods, we have discerned distinct patterns and variances among candidates and time periods. However, it's important to acknowledge the limitations of our study. First, we must contend with the blurred silhouettes that can affect our findings for Study 1 and data biases for both. Second, while we have delved into the realm of Instagram, a rich source of political content, we have yet to explore the types of content that elicit the highest levels of engagement. Future research can integrate additional datasets, including metadata from these posts, to unravel the content that garners the most engagement. Furthermore, studies could also examine how visual contents are disseminated across various platforms, such as Facebook and Twitter, to offer a comprehensive understanding of their impact on political discourse in digital settings.

**Funding**

This project was partially funded by the University of Amsterdam's RPA Human(e) AI and by the European Union's Horizon 2020 research and innovation programs No 951911 (AI4Media).